\documentclass[a4paper, 11pt, oneside]{article}

\usepackage[T1]{fontenc}
\usepackage [english]{babel}
\usepackage{natbib} 

\usepackage{amsmath,amssymb}
\usepackage{tabularx}
\usepackage{latexsym}
\usepackage{amsfonts}
\usepackage{mathtools} 
\usepackage{bm}
\usepackage{dsfont}
\usepackage{bbm}

\usepackage{amsthm}
\usepackage{amscd}
\usepackage[inline]{enumitem}

\usepackage{multirow}
\usepackage{graphicx,graphics,rotating}
\usepackage[graphicx]{realboxes} 
\usepackage{epstopdf} 
\usepackage[font=scriptsize,labelfont=bf,labelsep=period,textfont=it]{caption}

\usepackage{mathrsfs}
\usepackage[a4paper,top=2cm,bottom=2.5cm,left=2cm,right=2cm,height rounded,bindingoffset=0.5cm]{geometry}

\usepackage{etoolbox}
\usepackage{xpatch}  
\usepackage[gen]{eurosym}
\usepackage{lscape}
\usepackage[title,titletoc]{appendix}

\usepackage{sgame}
\usepackage{setspace}
\usepackage[font={small,sl}]{caption}
\usepackage{booktabs,caption} 
\usepackage[para,online,flushleft]{threeparttable}	

\usepackage{blindtext}
\usepackage{fancyhdr}
\usepackage{setspace}
\usepackage[explicit]{titlesec}
\usepackage[normalem]{ulem}
\usepackage{lipsum}
\usepackage{mwe}  
\usepackage{verbatim}
\usepackage{booktabs,array}  
\usepackage{listings}
\usepackage{subfig}
\usepackage[colorlinks=true,citecolor=blue,linkcolor = blue,urlcolor  = blue]{hyperref}
\usepackage{xcolor}

\usepackage{booktabs}
\usepackage[hang,flushmargin]{footmisc}

\usepackage{mathptmx}

\makeatletter

\def\citeapos#1{\citeauthor{#1}'s (\citeyear{#1})}

\title{\Large \bf Influence Analysis with Panel Data%
\footnote{
I thank Marcus Chambers, Marco Francesconi, Abhimanyu Gupta, Chiara Monfardini, and the participants to the $10^{th}$ ICEEE, $20^{th}$ German Stata Conference, $28^{th}$ IPDC, 2023 Petralia Workshop, and $29^{th}$ UK Stata Conference for the constructive suggestions. I am also grateful to Omar Hussein for his insightful comments and help.
This project was supported by the South East Network for Social Sciences (SeNSS) as part of the Doctoral Training Partnership (award ES/P00072X/1/2128216). This research was also funded by the UK Economic and Social Research Council award ES/S012486/1 (MiSoC).
The \textsc{Stata} commands for implementing the methodology presented in the paper are \texttt{xtlvr2plot} for unit-wise leverage-vs-residual plots, and \texttt{xtinfluence} for the influence analysis with panel data. The community-contributed are available on \href{https://github.com/POLSEAN/Influence-Analysis}{\texttt{https://github.com/POLSEAN/Influence-Analysis}} at the time of writing. %
}}
\author{Annalivia Polselli\thanks{Institute of Analytics and Data Science (IADS) and Centre for Micro-Social Change (MiSoc), University of Essex. Email: \href{annalivia.polselli@essex.ac.uk}{annalivia.polselli@essex.ac.uk}.}}
\date{Current version: \today}

\newcommand{\distas}[1]{\mathbin{\overset{#1}{\kern\z@\sim}}}%
\newsavebox{\mybox}\newsavebox{\mysim}
\newcommand{\distras}[1]{%
  \savebox{\mybox}{\hbox{\kern3pt$\scriptstyle#1$\kern3pt}}%
  \savebox{\mysim}{\hbox{$\sim$}}%
  \mathbin{\overset{#1}{\kern\z@\resizebox{\wd\mybox}{\ht\mysim}{$\sim$}}}%
}

\newcommand{\bfe}{\widehat{\bm{\beta}}}

\newcommand{\bi}{\widehat{\bm{\beta}}_{(i)}}
\newcommand{\bij}{\widehat{\bm{\beta}}_{(i,j)}}
\newcommand{\bj}{\widehat{\bm{\beta}}_{(j)}}

\newcommand{\bbeta}{\bm{\beta}}
\newcommand{\x}{\mathbf{x}_{it}}

\newcommand{\xit}{\widetilde{\mathbf{x}}_{it}}
\newcommand{\xis}{\widetilde{\mathbf{x}}_{is}}

\newcommand{\xii}{\mathbf{X}_i}

\newcommand{\Xii}{\widetilde{\mathbf{X}}_i}
\newcommand{\Xjj}{\widetilde{\mathbf{X}}_j}
\newcommand{\Xll}{\widetilde{\mathbf{X}}_l}
\newcommand{\Xj}{\widetilde{\mathbf{X}}_{i(j)}}
\newcommand{\Xij}{\widetilde{\mathbf{X}}_{ij}}
\newcommand{\X}{\widetilde{\mathbf{X}}}

\newcommand{\SSXX}{\mathbf{S}_{XX}}

\newcommand{\Sxx}{\mathbf{S}_N}

\newcommand{\bVn}{\overline{\mathbf{V}}_N}

\newcommand{\Vi}{\mathbf{V}_i}

\newcommand{\y}{y_{it}}
\newcommand{\yii}{\mathbf{y}_i}

\newcommand{\yi}{\widetilde{\mathbf{y}}_i}
\newcommand{\yj}{\widetilde{\mathbf{y}}_j}
\newcommand{\yij}{\widetilde{\mathbf{y}}_{ij}}
\newcommand{\Y}{\widetilde{\mathbf{Y}}}

\newcommand{\hitt}{h_{ii,tt}}
\newcommand{\hits}{h_{ii,ts}}

\newcommand{\Hi}{\mathbf{H}_{i}}

\newcommand{\Hj}{\mathbf{H}_j}
\newcommand{\Hij}{\mathbf{H}_{ij}}
\newcommand{\Hji}{\mathbf{H}_{ji}}
\newcommand{\blockH}{\mathbf{\overline{H}}_{ij}}

\newcommand{\Mi}{\mathbf{M}_i}
\newcommand{\Mj}{\mathbf{M}_j}

\newcommand{\blockM}{\mathbf{\overline{M}}_{ij}}

\newcommand{\error}{u_{it}}
\newcommand{\errori}{\mathbf{u}_i}

\newcommand{\uii}{\widetilde{\mathbf{u}}_i}

\newcommand{\ull}{\widetilde{\mathbf{u}}_l}

\newcommand{\resi}{\widehat{\mathbf{u}}_i}
\newcommand{\resj}{\widehat{\mathbf{u}}_j}
\newcommand{\resl}{\widehat{\mathbf{u}}_l}
\newcommand{\resij}{\widehat{\mathbf{u}}_{ij}}

\newcommand{\A}{\mathrm{A}}

\newcommand{\B}{\mathrm{B}}

\newcommand{\C}{\mathrm{C}}
\newcommand{\Ci}{\mathrm{C}_{ii}(\bfe)}

\newcommand{\Cij}{\mathrm{C}_{ij}(\bfe)}

\newcommand{\cCij}{\mathrm{C}_{i(j)}(\bfe)}

\newcommand{\D}{\mathrm{D}}

\newcommand{\Kij}{\mathrm{K}_{j|i}}
\newcommand{\MSij}{\mathrm{M}_{i(j)}}
\newcommand{\F}{\mathrm{F}}

\newcommand{\iotat}{\bm{\iota}}
\newcommand{\Vv}{\mathbf{V}}

\newcommand{\inv}{^{-1}}
\newcommand{\invsqrt}{^{-1/2}}

\newcommand{\N}{\mathcal{N}}
\newcommand{\uniform}{\mathcal{U}}

\newcommand{\E}{\mathbb{E}}

\newcommand{\Sgma}{\bm{\Sigma}_i}

\newcommand{\Ssgma}{\bm{\Sigma}}

\newcommand{\bSgma}{\overline{\bm{\Sigma}}_N}

\newcommand{\bb}{\big\|}
\newcommand{\Bb}{\Big\|}

\newcommand{\I}{\mathbf{I}}
\newcommand{\Ii}{\mathbf{I}_i}
\newcommand{\Ij}{\mathbf{I}_j}
\newcommand{\Iij}{\mathbf{I}_{ij}}

\newcommand{\zero}{\bm{0}}

\begin{document}
\maketitle

 \begin{abstract}
The presence of units with extreme values in the dependent and/or independent variables (i.e., vertical outliers, leveraged data) has the potential to severely bias regression coefficients and/or standard errors. This is common with short panel data because the researcher cannot advocate asymptotic theory. Example include cross-country studies, skill-cell analyses, and experimental studies. Available diagnostic tools may fail to properly detect these anomalies, because they are not designed for panel data. In this paper, we formalise statistical measures for panel data models with fixed effects to quantify the degree of leverage and outlyingness of units, and the joint and conditional influences of pairs of units.  We first develop a method to visually detect anomalous units in a panel data set, and identify their type. Second, we investigate the effect of these units on LS estimates, and on other units’ influence on the estimated parameters. To illustrate and validate the proposed method, we use a synthetic data set contaminated with different types of anomalous units. We also provide an empirical example.  \\ 

\end{abstract}

\section{Introduction}

Observational data often contain data points that exhibit extreme values in the response-space (vertical outliers, VO), in the covariate-space (good leverage points, GL), or in both directions (bad leverage points, BL) \citep{rousseeuw1990,silva2001}. 
The presence of these anomalies creates bias in the least square (LS) estimates (i.e., coefficients and/or  standard errors), and can potentially invalidate the results because econometric techniques based on the mean are more sensitive to extreme realisations \citep{donald1993,bramati2007,verardi2009}. Their impact is more severe with panel data because these anomalies may be carried over the time series of a unit, exacerbating the bias \citep{bramati2007}. This is particularly problematic in short panel data sets (few cross-sectional units $N$ are observed over multiple time periods $T$) because the small cross-sectional size does not allow to mobilise the benefits of the asymptotic theory. These data are common in some fields of economics due to the structure of the data or the nature of the study. Some examples are: worldwide country-level macro-data (e.g., 50 United States, 27 European countries); skill-cell data sets (e.g., with education-experience-state cells as unit of observation); or experimental data, where it is costly for the researchers to recruit many participants to follow in successive waves.  



Popular tools for the detection of such data points include diagnostic plots (e.g., leverage-vs-residual plots), and  measures of influence (e.g., \citeapos{cook1979} distance). These respectively serve to identify the type of anomaly, and assess the individual influence on the LS estimates by removing one unit at a time.  Despite the ease of their interpretability and computation, these measures may fail to detect anomalous units when surrounded by other cases as they do not assess mutual influence of units. This phenomenon -- known as the \emph{masking effect} -- is documented for Cook-like measures \citep{atkinson1985, chatterjee1988,rousseeuw1990,rousseeuw1991}, but can be overcome by measures that delete two cases at a time from the sample, such as,  joint and conditional measures for the influence \citep{lawrance1995}. 
A visual inspection of these anomalies may be complicated with many variables \citep{rousseeuw1991,bramati2007}, thus the need of a method that identifies these anomalies based on statistical measures.

In this paper, we develop a method to: (i) systematically detect anomalous units in a panel data set, and identify their type with leverage-vs-residual plots; and (ii) investigate the effect of these units on LS estimates, and on other units’ influence on the estimated parameters with influence plots. For this purpose, we first formalise statistical measures -- i.e., individual leverage and normalised residual squared -- that quantify the degree of leverage and outlyingness of units all over its time series. These are then used to produce `unit-wise' leverage-vs-residual plots.
Second, we  build on \citeapos{lawrance1995} pair-wise approach by proposing measures for joint and conditional influence of units $i$ and $j$ that are designed for panel data models with fixed effects. 
We illustrate the method with synthetic data that have been contaminated with anomalous units. We then apply our method to real data. 

We show that our method detects problematic units that will not otherwise be detected  by the conventional methods. Specifically, leverage-vs-residual plots provide an initial picture of the anomalies in the data, being more informative about the existence and type of anomaly. Then, influence plots complete the analysis by displaying the strength and direction of the connection between units $i$ and $j$, showing how the influential units affect the contribution of other units on the LS estimates. 
The strength of this method is that a unit, which is not individually influential according to Cook’s distance, will be detected if it is influential either jointly with another unit, or in the absence of another highly influential unit.  The method can help the researcher understand how some units may be driving most of the effect on LS estimates  \emph{after} the main regression analysis, or explore the data set and identify potential anomalies \emph{before} any regression analysis. 

Once a unit is detected as potentially influential, the researcher should not proceed with its deletion because it might be a valid observation that should be properly addressed. For example, robust estimation accounts for anomalies that affect the estimated coefficients (i.e., VO and BL) \citep{bramati2007,verardi2009,aquaro2013,aquaro2014,Jiao2022}, and jackknife-type standard errors deal with anomalies that affect the statistical inference (i.e., GL) \citep{mackinnon1985,chesher1987,davidson1993,mackinnon2013,belotti2020,polselli2022,mackinnon2023cluster,mackinnon2023}.\footnote{These methods will not be discussed further in this study.}

This paper contributes to the literature on diagnostic and influential measures as follows.
While Cook-like distances are still on of the most popular diagnostic measures for the detection of individually influential observations,\footnote{Several variants of \citeapos{cook1979} measure have been proposed for both cross-sectional and panel data models. In the cross-sectional framework, \citet{martin2009} extend the measure to logistic models, \citet{pinho2015} to generalized linear mixed models, and \citet{martin2015} to log-linear models. \citet{banerjee1997} are the first to formalise a leave-one-out diagnostic measure to detect outliers and influential points for panel data, specifically in linear longitudinal models with random effects. \citet{belotti2020} have extended \citeapos{banerjee1997} distance to linear panel data models with fixed effects.} authors have not yet addressed how to overcome the \emph{masking effect}. We build on \citeapos{lawrance1995} pair-wise measures to investigate the reciprocal influence of units. We present an adaptation of the joint and conditional influence measures for linear panel data models with fixed effects. 
With this measures, we can detect units that would have not been detected otherwise by Cook's distance because not individually influential.
Because the joint and conditional measures resemble a directed and weighted adjacency list from network analysis,\footnote{In a directed graph, the effect of unit $i$ on unit $j$ differs from $j$ to $i$; in a weighted graph, the intensity of the effect of each unit differs.} we propose a graphical representation of these measures that plots the influence of unit $j$ on $i$'s influence such as to show the existence and the strength of the connections between pairs of units. Using scatter plots or heat plots facilitates the visual detection and classification of anomalous data points, and their influence on other units. This representation is innovative in the literature on diagnostic measures, where \citeapos{atkinson1993} stalactite plot pioneers the visual inspection of multivariate outliers in the data. 

This paper is closely related to \citet{mackinnon2023cluster,mackinnon2023}. They derive a measure for the overall leverage of clusters for cross-sectional data.  We follow their derivation to calculate the overall leverage of a unit all over its history. Because we have repeated measures of the variables for each unit, our formulation addresses the presence of fixed effects by transforming the variables with the within-group (WG) transformation. 


The rest of the paper is structured as follows. In Section~\ref{sec:out_lev}, we start by defining the three types of anomalies. In Section~\ref{sec:diagnostics}, we first introduce the panel data models and estimators, and then present the measures of influence for panel data. Section~\ref{sec:method} discusses the proposed method using a synthetic data set. Section~\ref{sec:examples} applies the method to an empirical exercise. Section~\ref{sec:conclusion} concludes.


 
\section{Anomalous Units}\label{sec:out_lev}
Longitudinal data are extremely likely to contain units with values of the dependent and/or independent variables that follow a different pattern from the main cloud of the data in the response- and/or covariate-spaces.  These are classifiable as \emph{vertical outliers} (VO) if the extreme values are in the outcome variable, \emph{good leverage} (GL) points if the extreme values are in the covariates, or \emph{bad leverage} (BL) points if they are in both directions. 

\begin{figure}[th!]
    \caption{Example of anomalous units}\label{fig:anomalies}
    \centering
    \includegraphics[scale=.95]{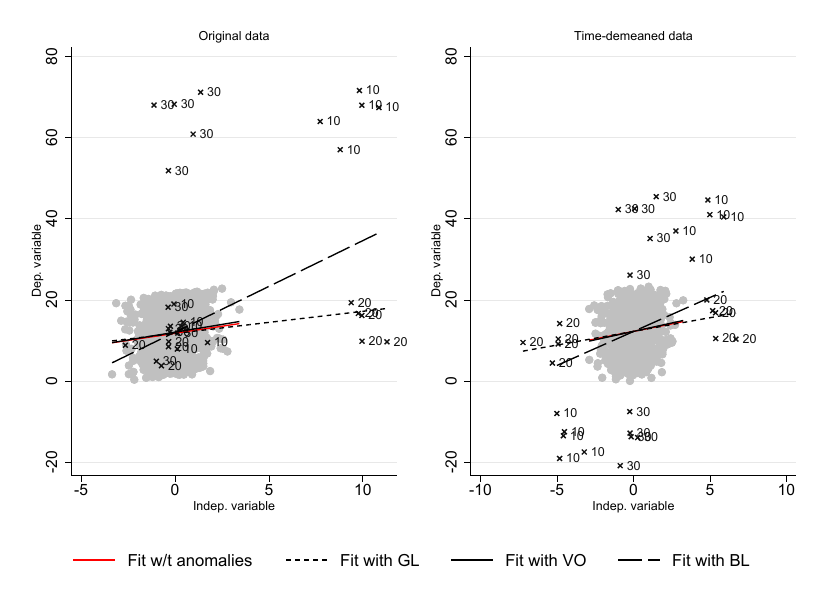}
    \begin{minipage}{.9\linewidth}
    \vspace{2mm}
    \small
    \emph{Note: The graphs show the relationship between the outcome variable and the only regressor. The original data is contaminated with three anomalous units: 10 is a bad leverage unit, 20 is a good leverage unit, and 30 is a vertical outlier. From left to right: scatter plot of original data, scatter plot data after the within-group transformation. The red dotted line is fitted using uncontaminated units only; the dashed line using uncontaminated and good leverage points; the dash-dot line using uncontaminated and bad leverage points; the solid line uncontaminated units and vertical outliers.}
\end{minipage}
\end{figure}

Vertical outliers (unit 30 in Figure~\ref{fig:anomalies}) are anomalous in the response-factor space, follow the opposite trend of the main cloud of data points, and lay far from the regression line as if they were generated from a different process \citep[p.141]{chatterjee1986, greene2012}. They display large squared normalised residual, and their presence alters the estimated LS intercept (here, the fitted line shifts upwards), undermining the accuracy of the estimator but not its precision \citep{verardi2009}. 

Leveraged data points appear isolated from the rest of the data but, unlike VO, follow the same trend of the rest of the data \citep[p.140]{chatterjee1986, greene2012}. 
Leverage points can be distinguished in `good' or `bad' as follows. The former (unit 20 in Figure~\ref{fig:anomalies}) exhibit unusually extreme values in the covariates and lie on the predicted regression line enhancing the precision of the regression fit; whereas, the latter (unit 10 in Figure~\ref{fig:anomalies}) possess extreme values in both input and response direction, lie far from the plane where the bulk of data points are and are not fitted by the regression model \citep{rousseeuw1991,bramati2007}. 
While BL realisations adversely affect the estimated LS coefficients (both the intercept and the slope of the regression), GL points add variability in the sample allowing for a better fit of the data at the cost of a deteriorated statistical inference \citep{mackinnon1985,chesher1987,silva2001,verardi2009}.

In a panel data sets, anomalous realisations in the time series of a unit may appear either as isolated cases in the time-series of different units (\emph{cell-concentrated} points), or concentrated in the time-series of the same units (\emph{block-concentrated} points) \citep{bramati2007}. 


\section{Measures of Influence in Panel Data Models}\label{sec:diagnostics}
Let $\{\y,\x\}_{t=1}^T$ be repeated measurements of unit $i \in \mathcal{I}=\{1,\dots,N\}$ over $t$ time periods, and that $T_i=T$ for all $i$ (i.e., balanced panel).\footnote{The data is assumed to be balanced to simplify the notation, but it can be equally applied to unbalanced panel data with appropriate change in notation.} Consider a linear panel regression model with unobserved individual heterogeneity as follows
\begin{equation}\label{eqn:fe_it}  
\y =  \x' \bm{\beta} + \alpha_i + \error, 
\end{equation}
\noindent where $\y$ is the response variable for the cross-sectional unit $i$ at time period $t$; $\x$ is a $k\times1$ vector of time-varying inputs, $\bm{\beta}$ is a $k\times1$ vector of parameters of interest; $\alpha_i$ is the individual-specific unobserved heterogeneity (or \emph{fixed effects}); and $\error$ is a stochastic error component. 

Stacking observations over time $t$, model \eqref{eqn:fe_it} becomes
 \begin{equation}\label{eqn:fe_i}   
\yii =   \xii\bm{\beta} +\bm{\alpha}_i+ \errori,  \,\,\, \text{for all}  \,\,\, i= 1,\dots, N, 
\end{equation}
\noindent where $\yii$ is a $T\times1$ vector of the dependent variable; $\xii$ is a $T\times k$ matrix of time-varying independent variables; $\bm{\alpha}_i=\alpha_i\iotat$ is a $T\times1$ vector of unobserved individual fixed effects, and $\iotat$ is a vector of ones of order $T$; and $\errori$ is a $T\times1$ vector of one-way error component. 

Model~\eqref{eqn:fe_i} can be consistently estimated using OLS on the time-demeaned model
 \begin{equation}\label{eqn:fewg}   
\yi =  \Xii\bm{\beta} + \uii, \,\,\, \text{for all}  \,\,\, i= 1,\dots, N,  
\end{equation}
\noindent where $\yi=(\I_T-\, T\inv\iotat\iotat')\yii$ is $T\times1$; $\Xii=(\I_T-\, T\inv\iotat\iotat')\xii$ is $T\times k$; and $\uii=(\I_T-\, T\inv\iotat\iotat')\errori$ is $T\times1$. Note that $(\I_T-\, T\inv\iotat\iotat')\bm{\alpha}_i=\zero$ as $T\inv \iotat\iotat' \bm{\alpha}_i=\bm{\alpha}_i$. 
The OLS estimator of \eqref{eqn:fewg} is the well-known \emph{within-group} estimator of the true population parameter with formula 
\begin{equation}\label{eqn:b_wg}  
 \bfe= \Biggl( \sum_{i=1}^N \Xii'\Xii \Biggr)\inv \sum_{i=1}^N  \Xii'\yi.   
\end{equation}
The estimator  is unbiased and asymptotically normally distributed under conventional model assumptions.

The \emph{within-group} estimator~\eqref{eqn:b_wg}  without the whole history of unit $i$ is the  Leave-One-Out (L1O) estimator\footnote{The L1O estimator for panel data models is derived in \citet{banerjee1997} for mixed effects models, and \citet{belotti2020} for fixed effects models. We show the asymptotic distribution of $\bi$ in Appendix~\ref{sec:distr_diagnostic}.} as follows
 \begin{equation}\label{eqn:l1o_sec}
\bi= \bfe- \bigl(\X'\X\bigr)\inv \Xii' \Mi\inv\resi
  \end{equation}
 \noindent  where $\Mi\inv=(\I_T-\Hi)\inv$ with  $\Hi=\Xii (\X'\X)\inv \Xii'$, and $\resi =\yi - \Xii\bfe$ is the residual term. The L1O estimator estimates the effect of the covariates on the outcome variable by removing one unit at a time from the sample. This is a general measure of the influence exerted by that unit on the estimated coefficients or on the model fit.

The \emph{within-group} estimator~\eqref{eqn:b_wg} without the full history of pairs of units $\{i,j\}$ is the Leave-Two-Out (L2O) estimator\footnote{The derivation of Formula~\eqref{eqn:l2o_sec} and its distribution in Appendices~\ref{app:l2o} and~\ref{sec:distr_diagnostic}.} below
 \begin{equation}\label{eqn:l2o_sec}
\bij= \bi - \bigl(\X'\X\bigr)\inv  \bigl(\Xii'\Mi\inv\Hij+\Xjj'\bigr)\bigl(\Mj-\Hij'\Mi\inv\Hij\bigr)\inv \bigl(\Hij' \Mi\inv\resi+\resj\bigr)
  \end{equation}
\noindent where $\Mj = \Ij-\Hj$ with $\Hij = \Xii (\X'\X)\inv \Xjj' $, and $ \Hj = \Xjj (\X'\X)\inv \Xjj' $ noting that \mbox{$\Hji = \Hij'$}. The deletion of pairs of units one at a time from the sample captures the influence exerted by that pair on the estimated parameters or on the model fit.


\subsection{Leverage and Normalised Residual Squared}\label{sec:lvr2res}
The change in the magnitude of the estimated coefficients or the model fit after the deletion of a unit from the sample is informative of the influence of that unit. This can be explained by high residuals, high individual leverage, or both. In this section, we construct measures for panel data that assess the individual influence of a unit along these two dimensions. We define the  individual leverage and the normalised residual squared for panel data, that considers the full time series of a unit (`unit-wise' approach). 

The leverage of a unit measures the distance of the values of the covariates of a unit from those of other units in the sample. This information is provided by the leverage matrix of unit $i$, $\Hi =\Xii \bigl(\X'\X\bigr)\inv\Xii'$, which is a $T\times T$ positive-definite and symmetric with diagonal elements $\hitt = \xit'(\X'\X)\inv\xit$, and off-diagonal elements  $\hits = \xit'(\X'\X)\inv\xis$.\footnote{With only one time period (cross-sectional data), the leverage matrix of unit $i$ is the scalar $h_i = \mathbf{x}_i'(\X'\X)\inv\mathbf{x}_i$ with $\mathbf{x}_i$ a $k\times1$ vector of covariates and the constant term.} 


We calculate the nuclear (or trace) norm of $\Hi$ to summarise the leverage exerted by each unit over their time series, as in \citet{mackinnon2023cluster,mackinnon2023} for clustered data.\footnote{Similarly, \citet{mackinnon2023} calculate the leverage matrix of cluster $G$ populated with $N_g$ units. They use the nuclear norm operator to provide an overall assessment of the influence of each group in a way that is computationally feasible. We face a similar challenge with panel data, where the unit can be seen as the $N$-th cluster populated with a total of $T$ time realisations. Therefore, their notion of leverage for cluster-level data is comparable to some extent to ours for panel data with some adjustments (i.e., the WG transformation to remove the fixed effects).} The overall leverage of unit $i$ can be calculated as $L_i = \mathrm{tr}\big(\Xii'\Xii \bigl(\X'\X\bigr)\inv\big)$. The average measure of the leverage exerted by all units in the sample is $ \frac{1}{N}\sum_{i=1}^N \mathrm{tr}(\Hi)= k/N$ by rearranging terms in the trace operator, noting that $\sum_{i=1}^N\Xii'\Xii=\X'\X$, and $k=\mathrm{rank}(\X)$. This result (similar to the cross-sectional case) can be used to establish the cut-off value for a unit to be considered as highly influential. For instance, a unit with an individual leverage that exceeds twice the value of the individual leverage in the sample, $L_i>2k/N$, is a high leverage unit. High leverage units possess unusually extreme values (too small or too large) in the regressors, and can be classified as GL unit.

We now define a measure to locate anomalous units in the $y$-space. We define the normalised residual squared for panel data as $\widehat{u}^*_{it} = \big(\widehat{u}_{it}/\sqrt{\sum_i\widehat{u}^2_{it}}\big)^2$, which informs on the outlyingness of unit $i$ at time $t$. We define $\widehat{\mathbf{u}}^*_i = [\widehat{u}^*_{i1}\, \dots\, \widehat{u}^*_{iT} ]'$ a $T\times1$ vector that contains the information of the influence exerted by unit $i$ in each time period $t\in\{1,\dots,T\}$ on the least squares estimates. 

We use the Euclidean norm of $\widehat{\mathbf{u}}^*_i$ to quantify the influence of unit $i$ in the output space such that $O_i= \sqrt{\sum_t  \widehat{u}^{*2}_{it}}$. 
The average normalised residual squared over the full sample of units is $N\inv\sum_{i=1}^N\widehat{u}^*_{it}= 1/N$ (like in the cross-sectional case). This value can be used as a reference to compare the levels of influence of each units in terms of the dependent variables. That is, a unit is said to be influential in the $y$-space if its value of  normalised residual squared exceeds twice the average value in the sample, i.e., $O_i> 2/N$.



\subsection{Measures for Joint and Conditional Influence }
Joint influence is the influence exerted by a pair $(i,j)$ on the LS estimates \emph{jointly}. It is obtained by comparing the estimated coefficients \emph{with} and \emph{without} full history of the pair in the sample. In formulae, 
\begin{equation}\label{eq:Cij_finite}
\Cij = \bigl(\bfe - \bij \bigr)'\Bigl(\X'\X\Bigr)\bigl(\bfe - \bij\bigr)(s^2K)\inv  \approx\F(\nu_1,\nu_2)
\end{equation}
\noindent where $N$ is the total number of cross-sectional units in the sample; $\nu_1=K$ is the total number of regressors including the constant, and $\nu_2 = (N-1)$ is the degrees of freedom at the denominator because of the clustering; $s^2=\nu_2\inv\sum_{i=1}^N\resi'\resi$ is the variance of the fitted model -- i.e., residual mean squared error (MSE) -- and is consistent for $\sigma^2$. $\Cij$ is symmetric for $i\ne j$.
When $i=j$, the measure informs on the individual influence of unit $i$  on the estimated coefficients for $\beta$. This can be interpreted as the Cook's distance for panel data \citep{banerjee1997,belotti2020}. Formula~\eqref{eq:Cij_finite} becomes
\begin{equation}\label{eq:Ci_finite}
\Ci = \bigl(\bfe-\bi\bigr)'\Bigl(\X'\X\Bigr)\bigl(\bfe-\bi\bigr)(s^2K)\inv \approx\F(\nu_1,\nu_2)
\end{equation}
$\Ci$ is hence a special case of the more general measure for the deletion of pairs of units, $\Cij$.

The joint effect is the ratio 
\begin{equation}
    \Kij = \frac{\Cij}{\Ci}
\end{equation}
\noindent and informs on unit $i$'s influence within the $(i,j)$ pair. For large values of this measure, unit $j$ alters the individual effect of unit $i$ on the estimated coefficients. In other words, the most influential unit in the pair $j$ contributes to drive most of the effect on LS estimates by \emph{enhancing} or \emph{reducing} the effect of the least influential $i$. The assessment can be done in conjunction with the conditional effect below. 

Because joint measures do not compare individual influences arising  \emph{before} and \emph{after} the deletion of another unit, the notion of conditional influence is needed. Conditional influence is the influence exerted by unit $i$ on the LS coefficients \emph{conditional on} removing unit $j$ from the sample. It shows how the absence of unit $j$ alters the influence of unit $i$ on the LS estimates. In formulae,
\begin{equation}\label{eq:cCij}
\cCij = \bigl(\bij-\bj\bigr)'\Biggl(\sum_{i \ne j}\Xj'\Xj\Biggr)\bigl(\bij-\bj\bigr)(s^2K)\inv  \approx\F(\nu_1,\nu_2) 
\end{equation}
\noindent where $\Xj$ is a $(N-1)\times K$  matrix of time-demeaned regressors without unit $j$. The value of $\cCij$ is zero for $i=j$, and is not symmetric for $i\ne j$. The diagnostic measure $\cCij$ is not  a test statistics but the knowledge of its empirical distribution can be used to extrapolate cut-off values to assess the conditional influence of units.

The conditional effect is the ratio 
\begin{equation}
    \MSij = \frac{\cCij}{\Ci}
\end{equation}
\noindent quantifying unit $i$'s influence \emph{before} and \emph{after} the deletion of unit $j$ from the sample. For $\MSij\ge1$, unit $j$ is said to \emph{mask} the influence of unit $i$ because the influence of $i$ increases without $j$ in the sample. When $\MSij<1$, unit $j$ is said to \emph{boost} the influence of unit $i$ because the influence of $i$ decreases without $j$. With this information, we can interpret the results of the joint effect as follows. When the influence of $i$ is masked (boosted) by $j$, $j$ is reducing (enhancing) the effect of $i$.


These measures follow a F-distribution with $\nu_1$ degrees of freedom at the numerator and $\nu_2$ degrees of freedom at the denominator.\footnote{The derivation is shown in Appendix~\ref{sec:distr_diagnostic}.} These two measures are not test statistics but the information on their empirical distributions can be used to select distributional cut-off values to identify influential units, as recommended practice in the cross-sectional framework \citep{cook1979,martin2009,martin2015}. Cut-off values can be extrapolated from the median of the inverse cumulative F distribution with $\nu_1$ degrees of freedom at the numerator and $\nu_2$ degrees of freedom at the denominator. Alternatively,  non-distributional cut-off values of 1 or $4/N$ can be chosen \citet{bollen1985}.

The joint and conditional measures are constructed in a way that resembles a weighted and directed adjacency list from network analysis, which displays the existence and the strength of the links between pairs of units $(i,j)$. 
In a directed graph, the effect of unit $i$ on unit $j$ differs from $j$ to $i$; in a weighted graph, the intensity of the effect of each unit is different.  This allows us to mobilise graphical tools from network analysis that plot unit $j$'s influence on $i$'s influence by means of two-way scatter plots, or heat plots.  

\section{The Method}\label{sec:method}
In this section, we present the method to systematically detect anomalous units with panel data. Motivated by the fact that visual inspection of the data becomes complicated with more than two covariates \citep{rousseeuw1991,bramati2007}, we propose a method based on statistical measures for panel data that facilitates the detection and classification of anomalous units with many covariates. The method consists of two steps: first, the detection of anomalous units by type; and, second, the influence analysis to determine their effect on the LS coefficients, and on other unit's the influence on the LS parameters. 

This method is mainly designed for short panels because the bias arising from the presence of anomalous units in the sample is higher as the benefits of the asymptotic theory cannot be advocated. It is also thought to identify anomalies within clusters. The method can be employed \emph{before} the regression analysis to identify possibly problematic units, or \emph{after} to understand what units drive most of the average effect in the LS estimates. 

It goes beyond the scope of this paper to examine how anomalous units should be treated once they are detected with the proposed method. The deletion of any anomalous unit from the sample is not always the best option because the econometric literature offers techniques to handle these anomalies properly. A strand of the literature focuses on developing robust estimators to consistently estimate the parameter of interest by handling VO and BL points \citep{bramati2007,verardi2009,aquaro2013,aquaro2014,Jiao2022}. Another focuses on the downward bias of the statistical inference in the presence of GL points, recommending the use of jackknife-type standard errors  \citep{mackinnon1985,chesher1987,davidson1993,mackinnon2013,belotti2020,polselli2022,mackinnon2023}.


\subsection{Detection of Anomalous Units}
The first step consists in detecting the presence of anomalous units, and identifying their type from their location in the leverage-residual space. A graph that plots the individual leverage over the normalised residual squared (i.e., leverage-versus-residual plot) informs on the presence and the type of anomalous unit in the sample based on their location in the plane.  

\begin{figure}[th!]
    \caption{Comparison of leverage-versus-residual scatter plots}
    \label{fig:xtlvr2plot}
    \centering
    \includegraphics[scale=.7]{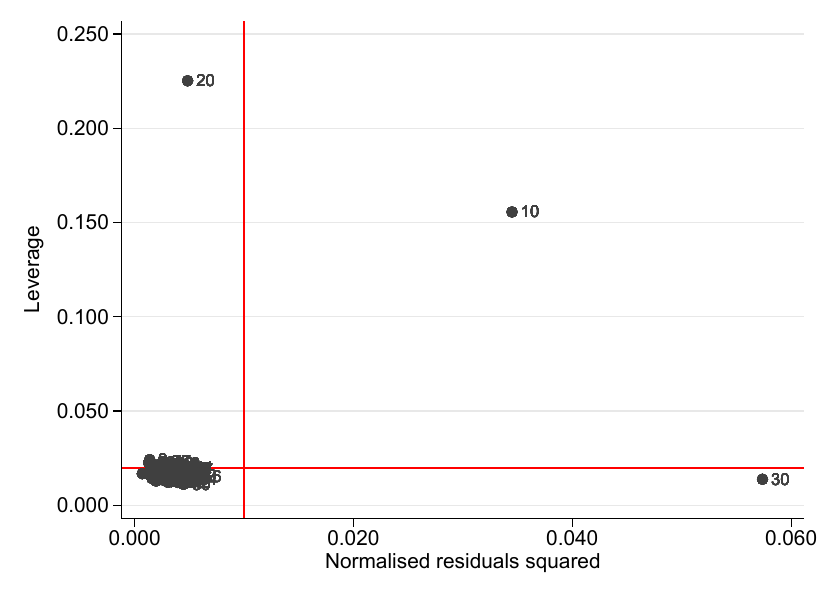}
    \begin{minipage}{.95\linewidth}
    \vspace{2mm}
    \small
    \emph{Note: The graph plots the leverage against normalised residual squared, evaluating the influence of each unit when its full time series is considered. The horizontal and vertical red lines mark the average values for high leverage and normalized residual squared, respectively. A unit can be considered anomalous if it exceeds twice the value of the individual leverage and/or the individual normalised residuals squared. The data is contaminated with three anomalous units: 10 is a bad leverage unit, 20 is a good leverage unit, and 30 is a vertical outlier.}
    \end{minipage}
\end{figure}

Figure~\ref{fig:xtlvr2plot} plots the individual leverage over the normalised squared residual.  The solid red lines mark the average values for individual leverage and normalised residuals squared. Cut-off values for high leverage and high residuals can be arbitrarily set as twice the average values. With this respect, plausible GL units possess high leverage but low squared residual and are located in the top-left quadrant; possible VO have high residual squared but low leverage and are located in the bottom-right quadrant; possible BL units display both high leverage and residual squared are in the top-right quadrant. Non-influential units are grouped in the cloud of points in the bottom-left quadrant or around the intersection of the two lines as they display low leverage and residual.

We illustrate the method to detect anomalous units in a leverage-residual plot with synthetic data that has been previously contaminated with anomalous units: unit 10 is a BL, unit 20 a GL, and unit 30 a VO. From the graph, the type of three units is correctly detected based on their leverage and residuals, and easily identifiable from their location on the plane. The rest of the units are not influential, as per data generating process, being confined to the bottom-left quadrant. 



\subsection{Influence Analysis}
Once anomalous units are detected and classified with the leverage-vs-residuals plot, the second part of the influence analysis shows how anomalies affect the LS estimates, and other units' influence on the estimated coefficients in terms of joint and conditional influence. Because the output resembles a weighted and directed adjacency matrix, we can mobilise graphical tools from network analysis (e.g., heat plots and scatter plots) that plot the measure of pair-wise influence of unit $j$ on unit $i$. In this example, we visualise the links and strength between pairs of units using heat plots.

\begin{figure}[t!]
    \caption{Influence analysis, heat plot}\label{fig:xtinfluece}
    \centering
    \includegraphics[scale=1.1]{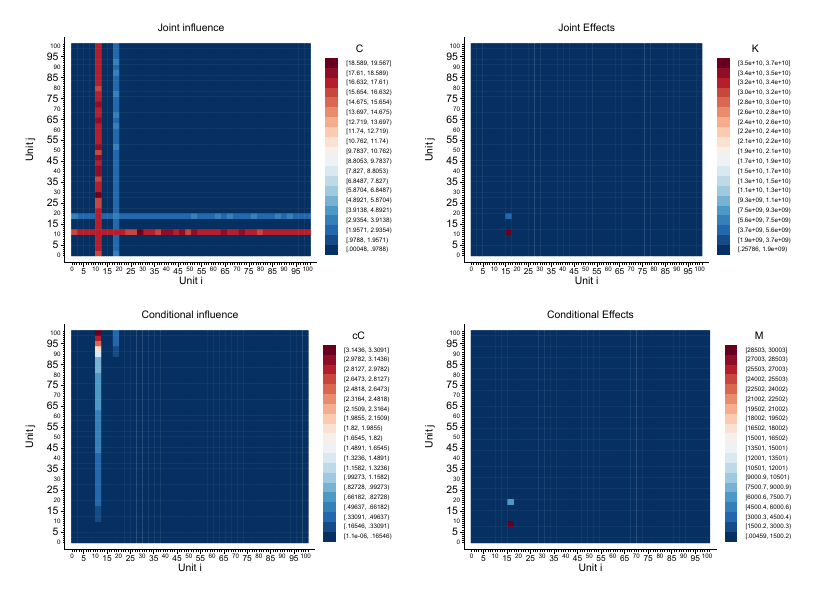}
        \begin{minipage}{.8\linewidth}
    \vspace{2mm}
    \small
    \emph{Note: The graphs plot the influence measures/effects for units $(i,j)$. The graph on the top-left displays the joint influence (or individual influence when $i=j$); the top-right graphs displays the joint effect; the bottom-left graph the conditional influence; the bottom-right graph the conditional effect. The data is contaminated with three anomalous units: 10 is a bad leverage unit, 20 is a good leverage unit, and 30 is a vertical outlier.}
\end{minipage}
\end{figure}

Figure~\ref{fig:xtinfluece} shows four plots with values of the joint influence, joint effects, conditional influence, and conditional effects  (respectively from top to bottom, left to right). The colour scale from dark blue to red shows the degree of influence/effect from the smallest to the largest value. The influence analysis consists of four steps as follows.

\textbf{1. The joint influence plot} displays the joint influence of pairs of units when $i\ne j$, and the individual influence when $i\ne j$. The diagonal elements represent Cook's distance for panel data. Units 10 and 20 -- generated as BL and GL units, respectively -- have high individual influence, largely exceeding the distributional cut-off 0.694 and the conventional Cook's Distance cutoff of 1. These units reciprocally have high joint influence in correspondence of another anomaly of the same type (e.g., with unit 30 as being a VO) and also with the rest of or almost all the units. 

\textbf{2. The joint effect plot} shows joint effects $\Kij$ of pairs of units. The values of this measure are extremely high for pairs of units ${i,j}$ equal to $\{17, 10\}$, $\{17, 20\}$. 
The $j$-th units that exert most of the influence in the pair are the BL and GL units that was already detected with the leverage-vs-residual plot and with the joint influence plot whereas the $i$-th unit was not. Detected units $j=\{10, 20\}$ \emph{alter} the individual effect of units $i=17$ by either \emph{reducing} or \emph{enhancing} its effect in the LS estimation. The direction of the effect (upwards or downwards) can be inferred from the outcome of the conditional effect in step 4.
 
\textbf{3. The conditional influence plot} highlights the influence of unit $i$ \emph{conditional on} removing $j$ from the sample. The conditional influence is quite high for already individually influential units (i.e., 10, and 20), by construction; these are the BL and GL units. This measures are informative for the construction of the conditional effects.

\textbf{4. The conditional effect plot} plots the measure $\MSij$ for the pair $\{i,j\}$. Units $j=\{10, 20\}$ are \emph{masking} the influence of unit $i=17$ as $\MSij\gg1$; no unit is \emph{boosting} the effect of any other. This information is useful to conclude  that the $j$-th units detected in step 2 are \emph{reducing} the effect of unit 17.  

\section{Empirical example}\label{sec:examples}
In this section, we illustrate an application of the method for the detection of anomalous units in panel data sets. We use the data and the regression equation  corresponding to Column~(2a) of Table~4 in \citet{berka2018}.\footnote{With this exercise, we do not intent to support or invalidate the original results.} The sample consists of 9 countries observed over the time period 1995--2007. The illustrative example is relevant to our case because of the small cross-sectional sample size, and the ease of interpretability of the regression output. 

\citet{berka2018} study the relationship between real exchange rate and sectoral productivity in nine Eurozone countries finding a strong correlation between productivity and real exchange rates among high-income countries with floating nominal exchange rates.  The estimating regression equation is as follows
\begin{equation}\label{eqn:berka}
RER_{it} = \beta TFP_{it} + \x' \bm{\gamma} + \alpha_i + \error,
\end{equation} 
\noindent where $RER_{it} $ is the log real exchange rate (expenditure-weighted) expressed as EU15 average relative to country $i$ (an increase is a depreciation) in period $t$; $TFP_{it}$  log of TFP level of traded relative to nontraded sector in EU12 relative to country i at time t; $\x$ are other covariates;  $\alpha_i$ is  the country fixed effects; and $\error$ is the error term. The coefficients in Equation~\eqref{eqn:berka} are estimated using OLS after the \emph{within-group} transformation.

\begin{figure}[h!]
    \caption{Leverage-vs-residual plots, \citet{berka2018}}
    \label{fig:xtlvr2plot_berka}
    \centering
    \includegraphics[scale=.85]{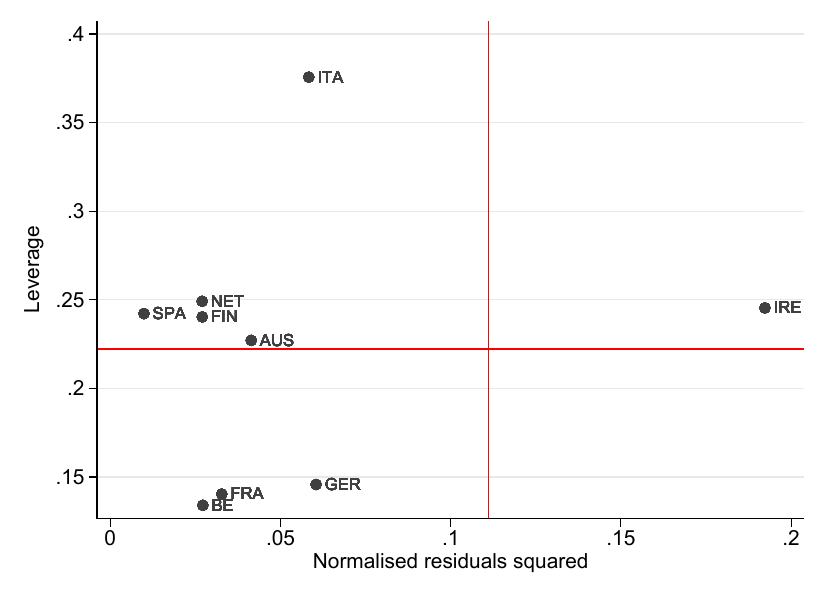}      
    \begin{minipage}{.95\linewidth}
    \vspace{2mm}
    \small
    \emph{Note: Data and regression equation  corresponds to Column~(2a) of Table~4 in \citet{berka2018}. From the graph, Italy can be a potential leverage point while Ireland a potential outlier.}
    \end{minipage}
\end{figure}

We are interested in identifying anomalous countries that may drive most of the effect on the LS estimate of $TFP$ based on Equation~\eqref{eqn:berka}. The leverage-vs-residual plot (Figure~\ref{fig:xtlvr2plot_berka}) highlights a group of six countries with individual leverage above the mean value -- where Italy may be a potential leverage units -- and only one country (Ireland) with the potential to be a BL or VO.

\begin{figure}[h!]
    \caption{Influence Analysis,  \citet{berka2018}}
    \label{fig:xtinfluence_berka}
    \centering
    \includegraphics[scale=1]{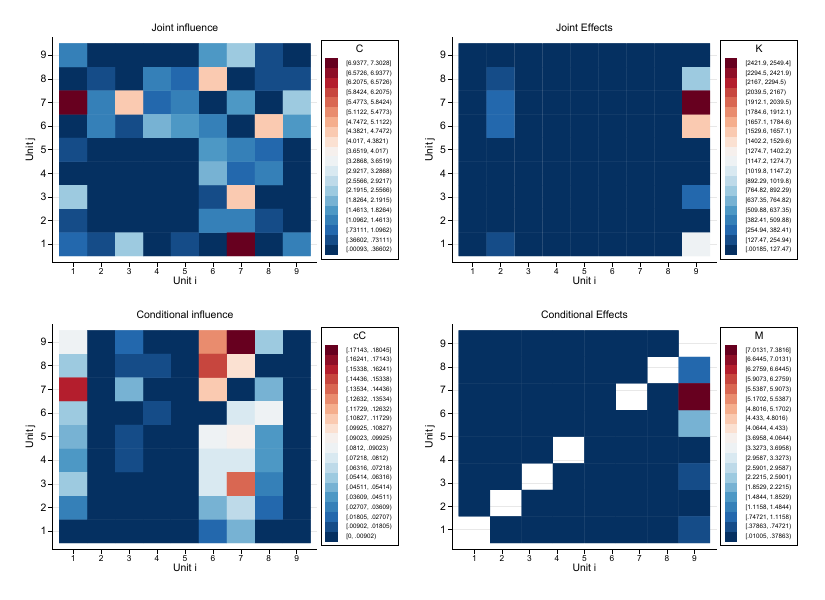}   
    \begin{minipage}{.95\linewidth}
    \vspace{2mm}
    \small
    \emph{Note: Data and regression equation  corresponds to Column~(2a) of Table~4 in \citet{berka2018}. The labels refer to 1-Austria, 2-Belgium, 3-Finland, 4-France, 5-Germany, 6-Ireland, 7-Italy, 8-The Netherlands, 9-Spain.  From the graph, the influence of Spain is masked and reduced by Ireland, Italy, and The Netherlands while is boosted and enhanced by Austria and Finland.}
    \end{minipage}
\end{figure}

Figure~\ref{fig:xtinfluence_berka}  shows the joint and conditional influence of pairs of countries. The labels refer to 1-Austria, 2-Belgium, 3-Finland, 4-France, 5-Germany, 6-Ireland, 7-Italy, 8-Netherlands, 9-Spain. On the main diagonal, individually influential countries  correspond to 6 (Ireland) and 7 (Italy). These exert high joint influence (red and pink squares) with other non-influential countries -- i.e., Ireland on Netherlands (unit 8), and Italy on Austria (unit 1) and Belgium (unit 2). Looking at the joint effects, units 1, 3, 6, 7, and 8 \emph{alter} the effect of unit 9 (Spain) on the LS estimates. The direction of the effect (upwards or downwards) can be inferred from the output of the conditional effect. Countries 6, 7, and 8 seems to be \emph{masking} the effect of country 9 whereas countries 1 and 3 are \emph{boosting} its contribution to the $TFP$ coefficient. Therefore, we can conclude by looking again at the joint effects plot that the former group of countries is \emph{reducing} the effect of Spain while the latter is \emph{enhancing} its influence.


\section{Conclusion}\label{sec:conclusion}
The presence of outliers and leveraged data points has the potential to bias regression coefficients and standard errors, which necessitates appropriate methods to detect and identify them.  Available measures -- such as, leverage-vs-residual plots and Cook-like measures -- may fail to correctly detect and classify them in panel data set. The method we propose overcomes this limitation by taking into account the panel structure of the data. We formalised the average leverage and average normalised residuals in a panel data setting to produce unit-wise leverage-residual plots. Then,  we developed two diagnostic measures for panel data models -- based on \citeapos{lawrance1995} cross-sectional measures -- and showed their statistical distributions. 

Overall, the method detected those units that exert more influence in the data set by taking into account the full history of a unit. The analysis of the joint and conditional effects  showed how their presence  alters the influence of other units in the sample and, hence, how their presence is driving the average effect on the LS estimates. The strength of this method is that a unit that is not individually influential according to Cook’s distance will always be detected if turns up to be influential \emph{jointly with} or \emph{conditional on} another highly influential unit. This will help the researcher understand how anomalous units drive their LS estimates. As a novelty, the individual and bilateral (joint and conditional) influence exerted by units in the sample are displayed with a network representation (e.g., heat plots and scatter plots).



The method can be used before the actual regression analysis to investigate the presence of anomalies in the sample, or after the regression analysis to understand what units are driving the estimated coefficients or standard errors.

Once anomalous units are properly detected and identified, the researcher should deal with their presence following the recommendations in the econometric literature. For example,  the literature recommends the use of robust estimators of the median (e.g., M-estimators, S-estimators, etc.)  in the presence of outliers \citep{bramati2007,verardi2009,aquaro2013,aquaro2014,Jiao2022}, whereas  heteroskedasticity-consistent standard errors based on jackknife methods in the presence of leveraged data \citep{hinkley1977,mackinnon1985,chesher1987,davidson1993,mackinnon2013,belotti2020,polselli2022,mackinnon2023cluster,mackinnon2023}.


\begin{spacing}{1}
\bibliographystyle{apalike}
\bibliography{99_main.bib}
\end{spacing}
\section{Figures}

\newpage\clearpage
\appendix
\section{Data Generating Process}\label{sec:app_dgp}
We use a synthetic data set to explain each step. The data generating process mainly follows \citeapos{bramati2007} simulation design with some differences. We generate a regression model as follows
\begin{align}
& \y = \beta_0+  \beta_1 x_{it}+ \alpha_i+ \epsilon_{it},  \text{ for all } i=1,\dots,N \text{ and } t=1,\dots,T \label{eqn:specification2} \\ 
&x_{it}\distas{} \N(0,1); \,\,\alpha_i\distas{} \uniform(0,20)  ; \,\, \epsilon_{it} \distas{} \N \bigl(0,1\bigr),  
\end{align}
\noindent where $\beta_0=1$ and $\beta_1=0.5$.  We set $N=100$ and $T=20$. Equation~\eqref{eqn:specification2} is estimated using the \emph{within-group} approach.

Because we are interested in the simultaneous influence exerted by multiple atypical observations in the sample, we generate a synthetic data set with two anomalies per type -- i.e., VO, GL and BL units.  VO (units 10 and 40) are obtained by adding values drawn from $\N(50,1)$ to the original dependent variable. GL units (units 20 and 50) are generated by replacing the original value of the regressor with values drawn from $\N(15,1)$. BL units  (units 30 and 60) are created  by adding $\N(50,1)$ to the original dependent variable, and replacing the original values of the independent variable with $\N(10,1)$. We generate \emph{block-concentrated} anomalies by contaminating $t\le10$ in the time series of units 10, 20, and 30, and $t\le5$ in the time series of units 40, 50, and 60.


\section{Derivation of L2O Estimator}\label{app:l2o}
Suppose that the panel is balanced with $T_i= T_j=T$.\footnote{This assumption simplifies the notation. However, the L2O estimator can also be estimated with unbalanced panel data, provided thatthe formula is modified to account for $T_i\ne T_j$.}  Let the FE estimator $\bbeta$ without units $i$ and $j$ be
\begin{align}\label{eqn:bij}
\bij & = \biggl(\X'\X-\Xii'\Xii-\Xjj'\Xjj\biggr)\inv \biggl(\X'\Y - \Xii' \yi- \Xjj' \yj\biggr) \notag\\
     & = \biggl(\X'\X-\Xij'\Xij\biggr)\inv \biggl(\X'\Y - \Xij' \yij\biggr),
\end{align}
\noindent where $\X$ is a ${NT\times k}$ matrix, $\widetilde{\mathbf{X}}_l$ for $l\in\{i,j\}$ is a $T\times k$ matrix, and $\Xij=\bigl[\Xii \hspace{3mm} \Xjj\bigr]'$ a ${2T\times k}$ matrix; $\Y$ is a ${NT\times 1}$ vector, $\widetilde{\mathbf{y}}_l$ for $l\in\{i,j\}$  is a ${T\times 1}$ vector,  $\yij=\bigl[\yi \hspace{3mm} \yj\bigr]'$ is a ${2T\times 1}$ vector. 

Apply the Woodbury's formula%
\footnote{The standard Woodbury's formula calculates the inverse of the sum of matrices as follows $(\A+\B\D\C)\inv=\A\inv - \A\inv\B \,(\D\inv+\C\A\inv\B)\inv \C\A\inv$. In this specific case, we have  $\A=\X'\X$, $\B=-\Xij'$, $ \C= \Xij$, and $\D = \Ii \oplus \Ij \equiv \Iij$ which is a $2T\times 2T$ block diagonal matrix with identity matrices $\Ii$ and $\Ij$ on the main diagonal, and off-diagonal block matrices of zeros of dimension $T\times T$.}
to Equation~\eqref{eqn:bij} such that the L2O can be expressed as
\begin{align}
\bij	& =\Bigr\{\bigl(\X'\X\bigr)\inv +\bigl(\X'\X\bigr)\inv \Xij' \bigl(\Iij- \blockH\bigr)\inv\Xij\bigl(\X'\X\bigr)\inv\Bigr\}\biggl(\X'\Y -\Xij' \yij\biggr)\notag\\
	& = \bfe- \bigl(\X'\X\bigr)\inv \Xij' \blockM\inv\resij\label{eqn:bij2},	
\end{align}
\noindent where %
    $\blockH = \Xij\bigl(\X'\X\bigr)\inv\Xij' = %
    \begin{bmatrix}
    \Hi& \Hij\\ 
    \Hij' & \Hj
    \end{bmatrix} 
    = %
    \begin{bmatrix}
    \Xii (\X'\X)\inv \Xii' & \Xii (\X'\X)\inv \Xjj' \\ 
    \Xjj (\X'\X)\inv \Xii' & \Xjj (\X'\X)\inv \Xjj' 
    \end{bmatrix} 
$, noting that $\Hji = \Hij'$; %
 $\blockM=(\Iij-\blockH)=%
    \begin{bmatrix}
    \Mi& -\Hij\\ 
    -\Hij' & \Mj
    \end{bmatrix}
    = %
    \begin{bmatrix}
    \Ii-\Hi& -\Hij\\ 
    -\Hij' & \Ij-\Hj
    \end{bmatrix}
$; and %
$ \resij \big[\resi \,\, \resj \big]$ is a $2t\times 1$ vector.

Expanding last block of Equation~\eqref{eqn:bij2} with the formula for the inverse of partitioned matrices\footnote{The formula is %
        \begin{equation*}
        \begin{bmatrix}
        A &B \\ 
        C& D
        \end{bmatrix}\inv
        = 
        \begin{bmatrix}
        A\inv +A\inv B(D-CA\inv B)\inv CA\inv &-A\inv B(D-CA\inv B)\inv \\ 
        -(D-CA\inv B)\inv CA\inv & (D-CA\inv B)\inv
        \end{bmatrix}
        \end{equation*}
provided that $A\inv$ exists and $(D-CA\inv B)$ is invertible.}, provided that $\Mi$ and $\bigl(\Mj-\Hij'\Mi\inv\Hij\bigr)$ are non-singular to be invertible Equation~\eqref{eqn:bij2} becomes
\begin{align}
 \bij & = \bfe -\bigl(\X'\X\bigr)\inv \Bigl(  \Xii'\Mi\inv \resi  + \bigl(\Xii'\Mi\inv\Hij+\Xjj'\bigr)\bigl(\Mj-\Hij'\Mi\inv\Hij\bigr)\inv \bigl(\Hij' \Mi\inv\resi+\resj\bigr)\Bigr)\notag\\
& = \bi - \bigl(\X'\X\bigr)\inv \bigl(\Xii'\Mi\inv\Hij+\Xjj'\bigr)\bigl(\Mj-\Hij'\Mi\inv\Hij\bigr)\inv \bigl(\Hij' \Mi\inv\resi+\resj\bigr)\label{eqn:bbij}
\end{align}
\noindent where $\bi = \bfe -\bigl(\X'\X\bigr)\inv \Xii'\Mi\inv \resi$ is the L1O estimator.

\section{Distribution of Influence Measures}\label{sec:distr_diagnostic}	
\subsection{Model assumptions for the convergence of the estimators}
Let $\{\yii,\xii\}_{t=1}^T$ be a sequence of independent and identically distributed (\emph{iid}) sequence of random variables, and $\{\errori\}_{t=1}^T$ be an independent but not identically distributed (\emph{inid}) sequence of error terms for all $i = 1, \dots, N$. Under this notation $T$ is the full set of time information, and the total number of observations in the sample is  given by $n = N\cdot T$ with balanced data sets. The length of the time period is fixed and smaller than the number of units $T \ll N$ to guarantee the conventional $N\to \infty$ asymptotics. 

Define $\widetilde{\mathbf{w}}_i = (\I_T-\, T\inv\iotat\iotat')\mathbf{w}_i$ the variable after the within-group (mean-centered) transformation. Then, we rely on the assumptions below for the convergence of the estimators
\begin{enumerate}[label=\sc{asm.}{\arabic*}, leftmargin=1.7\parindent,  rightmargin=.5\parindent]
\item \emph{Strict exogeneity}: \\
 $\E\bigl(\uii|\Xii\bigr)=\E(\widetilde{u}_{it}|\xit)=0$, for all $i= 1,\dots, N$ and $t=1,\dots,T$; \label{item:exogeneity}
\item \emph{Rank condition}: $\Sxx \equiv N\inv \sum_{i=1}^N \Xii'\Xii$ is a finite symmetric matrix with full column rank $k$. \label{item:rank}
\item \emph{Moment conditions}:  \label{item:mom} 
 	\begin{enumerate}[label={\roman*}.]		
 	 	\item  $\E\big\|\Xii'\Xii\big\|<\infty$ for  $\Xii'\Xii\in \mathbb{R}^{k\times k}$;  \label{item:mom_x2} 			
 	 	\item  $\mathrm{sup}_{i}\,\E\big\|\Xii' \uii\big\|^{2+\delta}<\infty$ for some $\delta>0$, $\forall i$ and $\Xii' \uii \in \mathbb{R}^k$, \label{item:mom_Adistr}	
	\end{enumerate}	
 \noindent where $\|\cdot\|$ denotes the Euclidean norm.

\item  \emph{Heteroskedasticity}: $\bSgma=N\inv \sum_{i=1}^N\Sgma\to\Ssgma$, where the  matrix of the heteroskedastic disturbances $\Sgma=\E\bigl(\uii\uii'|\Xii\bigr)= \mathrm{diag}\bigl\{\sigma_{it}^2\bigr\}$ is  symmetric  of dimension $T$, finite, positive definite, and diagonal. \label{item:het}

\item  \emph{Convergence of variance-covariance matrix}: \\
$\bVn=N\inv \sum_{i=1}^N \, \Vi\to\Vv$, where $\Vi=\E\bigl(\Xii'\Sgma\Xii\bigr)$ and $\Vv$ is a finite positive definite $k\times k$ matrix. \label{item:V} 
\end{enumerate}

\subsection{Joint Influence}\label{sec:joint}	
We first establish the asymptotic properties of $\bi$  and show its equivalence to $\bfe$ in the sense that 
\begin{equation*}
\sqrt{N}\bigl(\bi-\bbeta\bigr) =\sqrt{N} \bigl(\bfe-\bbeta\bigr) + o_p(1)
\end{equation*}
\noindent implying that  $\bi$ and $\bfe$ share the same asymptotic distribution. Note that $\bfe$ is a consistent estimator of $\bbeta$ under the least square assumptions in this paper.
We first show the consistency \mbox{of $\bi$}. Using the L1O formula, we know that
\begin{align}\label{eqn:bib}
\bi- \bbeta 
	& =  (\bfe - \bbeta)  -  \bigl(\X'\X\bigr)\inv \Xii' (\I_T - \Hi)\inv\bigl(\uii - \Xii(\bfe- \bbeta )\bigr) 
\end{align}
Using the Triangle Inequality,
\begin{equation}\label{eqn:bi-bbeta}
	\small
\bb(\bi- \bbeta)\bb \le\bb (\bfe - \bbeta)\bb  + \bigg\| \biggl(\frac{1}{N}\X'\X\biggr)\inv\bigg\| \,\frac{1}{N}\bb \Xii\bb \bb(\I_T - \Hi)\inv\bb \bigl(\bb\uii \bb+ \bb\Xii\bb \bb\bfe- \bbeta \bb\bigr)
\end{equation}
\noindent  is $ o_p(1)$ because $\bb \bfe- \bbeta\bb$ is $o_p(1)$; $\bigl(N\inv\X'\X\bigr)\inv = \SSXX^{-1} +o_p(1)$ by~\ref{item:rank}, \emph{WLLN} and \emph{Slutsky's theorem};
$\Xii$ and $\uii$ are $O_p(1)$ because random variables with finite moments by  \ref{item:mom} and, therefore, $N\inv\bb \Xii\bb = O_p(N^{-1})$; and 
\begin{equation} \label{eqn:Mi2} 
\bb(\I_T - \Hi)\inv\bb %
\le\bb\I_T\bb + \bb\Xii\bb^2 \bb\bigl(\X'\X-\Xii'\Xii\bigr)\inv  \bb\\
\end{equation}
\noindent is O(1) because the first term on the right-hand-side, $\sqrt{T}$, is O(1) without a remainder term, and the second component is bounded above by $o_p(1)$ random variable.

 As a result, $\bi$ is a consistent estimator of the true value of the parameter $\bbeta$ as $\bi = \bbeta + o_p(1)$ from~\eqref{eqn:bi-bbeta}. Therefore, removing one unit does not have an impact on the estimates of the true value of the parameter as the cross-sectional units increase to infinity.
\par  We now show that the estimators $\bi$ and $\bfe$ have the same asymptotic distribution by applying the Reverse Triangle Inequality to $\sqrt{N}\bigl(\bfe  - \bbeta \bigr) - \sqrt{N}\bigl(\bi  - \bbeta \bigr)$ such that
\begin{equation}\label{eqn:bbi}
\bb\sqrt{N}\bigl(\bfe  - \bbeta \bigr)  - \sqrt{N}\bigl(\bi  - \bbeta \bigr) \bb 
\le\biggl\|\biggl(\frac{1}{N}\X'\X\biggr)\inv\biggr\| \frac{1}{\sqrt{N}}\,\Bb\Xii\Bb \bb(\I_T - \Hi)\inv\bb\bigl(\bb\uii\bb +\bb\Xii\bb\bb\bfe-\bbeta\bb \bigr)\\
\end{equation}
\noindent is bounded above by a $o_p(1)$ random variable because the first component of~\eqref{eqn:bbi} is $\big(\SSXX\inv+ o_p(1)\big)$ as $N\inv \X'\X\overset{p}{\to}\E(\X'\X)= \SSXX>0$ by the Central Limit Theorem;  the second term is $,O_p(N^{1/r-1/2})=O_p(1)$ for $r\ge2$ under \ref{item:mom}.\ref{item:mom_x2}; the third component is $O(1)$ by \eqref{eqn:Mi2}; and the last quantity in parenthesis is $O_p(1)$. 

We can conclude that  $\bi$ and $\bfe$ share the same asymptotic distribution as
\begin{equation}\label{eqn:bi_convergence}
    \sqrt{N}\bigl(\bi  - \bbeta \bigr) = \sqrt{N}\bigl(\bfe  - \bbeta \bigr)   +o_p(1)
\end{equation}
\noindent where $\sqrt{N}\bigl(\bfe  - \bbeta \bigr) \overset{d}{\rightarrow} \N \bigl(\zero, \SSXX\inv \Ssgma \SSXX\inv\bigr)$. This result holds for all $i = 1, \dots, N$. 

Now, we find the distribution of $\bfe -\bi $ using Shwarz Inequality and Triangle Inequality
\begin{align}
\bb\bfe -\bi\bb & =\bb \bigl(\X'\X\bigr)\inv \Xii' (\I_T - \Hi)\inv\bigl(\uii - \Xii(\bfe- \bbeta )\bigr) \bb\notag\\
	& \le\bb N\bigl(\X'\X\bigr)\inv\bb N\inv\bb \Xii\bb \bb(\I_T - \Hi)\inv\bb \bigl(\bb\uii \bb+ \bb\Xii\bb \bb\bfe- \bbeta \bb\bigr) \label{eqn:loo}
\end{align}
\noindent is $o_p(1)$ because $\bb \bfe- \bbeta\bb\overset{p}{\to}0$; $\bigl(N\inv\X'\X\bigr)\inv = \SSXX\inv +o_p(1)$ by~\ref{item:rank}, \emph{WLLN} and \emph{Slutsky's theorem};
$\Xii$ and $\uii$ are $O_p(1)$ because random variables with finite moments by  \ref{item:mom} and, therefore, $N\inv\bb \Xii\bb = O_p(N^{-1})$ for $r\ge2$; and $\bb(\I_T - \Hi)\inv\bb $ is bounded above by $o_p(1)$ random variable as in \eqref{eqn:Mi2}.

\par We follow the same reasoning as above to derive the distribution of $\bij$. We re-write~\eqref{eqn:bbij} by adding and subtracting $\bbeta$ as follows
\begin{equation}\label{eqn:bijb}
 \bij - \bbeta = \bigl(\bi - \bbeta\bigr) -\bigl(\X'\X\bigr)\inv \bigl(\Xii'\Mi\inv\Hij+\Xjj'\bigr) 
  \bigl(\Mj-\Hij'\Mi\inv\Hij\bigr)\inv \bigl(\Hij' \Mi\inv\resi+\resj\bigr)
\end{equation}
\noindent  Using Triangle Inequality, we have that
\begin{align}
\small
\bb\bij- \bbeta\bb& \le\bb \bi - \bbeta\bb  \label{eqn:bijb_a}\\
&+ \bigg\| \biggl(\frac{1}{N}\X'\X\biggr)\inv\bigg\| \label{eqn:bijb_c}\\
& \hspace{5mm}\biggl(\frac{1}{N}\bb\Xii\bb\bb\Mi\inv\bb\bb\Hij\bb + \frac{1}{N}\bb\Xjj\bb\biggr) \label{eqn:bijb_d}\\
& \hspace{5mm} \bigg\|\bigl(\Mj-\Hij'\Mi\inv\Hij\bigr)\inv\bigg\| \label{eqn:bijb_e} \\
& \hspace{5mm} \bigl(\bb\Hij\bb \bb\Mi\inv\bb\bb\resi\bb+\bb\resj\bb\bigr)\label{eqn:bijb_f}
\end{align}
\noindent  is $o_p91)$ because the term on the right-hand-side of \eqref{eqn:bijb_a} is  $o_p(1)$ by result~\eqref{eqn:bi-bbeta},  \eqref{eqn:bijb_c} is $\SSXX\inv+o_p(1)$ as $N\to \infty$ and $T$ fixed, and the rest is $O_p(1)$. Specifically, \eqref{eqn:bijb_d} is $O_p(1)$ because of $N\inv\bb \Xll\bb = O_p(N^{-1})$ for $l\in \{i,j\}$, $\bb\Mi\inv \bb=O(1)$ by \eqref{eqn:Mi2}, and $\bb\Hij\bb= N\invsqrt\bb\Xii\bb \bb(N\inv\X'\X)\inv\bb  N\invsqrt\bb\Xjj\bb=o_p(1)$; component~\eqref{eqn:bijb_e} is $O_p(1)$ because $\bigl(\Mj-\Hij'\Mi\inv\Hij\bigr)\inv=\i_T + o_p(1)$ since $\Mj = \I_T -N\invsqrt  \Xjj (N\inv\X'\X)\inv N\invsqrt \Xjj' = \I_T + o_p(1)$; and, \eqref{eqn:bijb_f}  is $O_p(1)$ because $\bb\resl\bb=\bb\ull\bb + \bb\Xll\bb \bb(\bfe- \bbeta )\bb = O_p(1)$ by  \ref{item:mom}. 

\par We show that  the estimators $\bij$ and $\bfe$  have the same asymptotic distribution. Apply the Reverse Triangle Inequality to $\sqrt{N}\bigl(\bfe  - \bbeta \bigr)  - \sqrt{N}\bigl(\bij  - \bbeta \bigr)$ such that we have
\begin{align}
	\bb\sqrt{N}\bigl(\bfe  - \bbeta \bigr) & - \sqrt{N}\bigl(\bij  - \bbeta \bigr) \bb \notag\\
&	\le\biggl\|\biggl(\frac{1}{N}\X'\X\biggr)\inv\biggr\| \biggl\|\frac{1}{\sqrt{N}}\, \Xii' \Mi\inv\resi \biggr\| \label{eqn:bbij_a}\\
&+\bigg\|\biggl( \frac{1}{N} \X'\X \biggr)\inv\bigg\| \bigg\|\frac{1}{\sqrt{N}}\biggl(\Xii'\Mi\inv\Hij+\Xjj' \biggr)\bigg\| \label{eqn:bbij_b}\\
&\hspace{5mm} \bigg\|\bigl(\Mj-\Hij'\Mi\inv\Hij\bigr)\inv\bigg\| \label{eqn:bbij_c}\\
&\hspace{5mm} \bigg\|\Hij' \Mi\inv\resi+\resj\bigg\| \label{eqn:bbij_d}\\
&= o_p(1)\notag
\end{align}
\noindent is $o_p(1)$ because \eqref{eqn:bbij_a} is $o_p(1)$ by result~\eqref{eqn:bbi}, and the overall quantity in~\eqref{eqn:bbij_b} is $o_p(1)$, since the first component is  $\SSXX\inv+o_p(1)$ and  multiplies a quantity that is $O_p(1)$ as in~\eqref{eqn:bijb_c}, noting that $N\invsqrt\bb \Xii\bb = O_p(N^{1/r-1/2})$ and, hence,  $O_p(1)$ with $r\ge2$; \eqref{eqn:bbij_c} is $O_p(1)$ as in \eqref{eqn:bi-bbeta}; \eqref{eqn:bbij_d} is $O_p(1)$  as  in~ \eqref{eqn:bijb_f}.  Therefore,  $\bij$ and $\bfe$ share the same asymptotic distribution.

\noindent Last step consists in deriving the joint distribution of $\bfe$ and $\bij$. Rearranging Equation~\eqref{eqn:bbij}, we obtain
\begin{multline}\label{eqn:b_bij}
  \bfe -\bij =\bigl(\X'\X\bigr)\inv  \Xii'\Mi\inv \resi  \\
 \hspace{-10mm} +\bigl(\X'\X\bigr)\inv \bigl(\Xii'\Mi\inv\Hij+\Xjj'\bigr)\\
 \bigl(\Mj-\Hij'\Mi\inv\Hij\bigr)\inv \bigl(\Hij' \Mi\inv\resi+\resj\bigr)
\end{multline}
\noindent Using Shwarz Inequality and Triangle Inequality
\begin{align}
 \bb \bfe -\bij \bb 
&	\le\biggl\|\biggl(\frac{1}{N}\X'\X\biggr)\inv\biggr\| \biggl\|\frac{1}{N}\, \Xii' \Mi\inv\resi \biggr\| \label{eqn:betabij_a}\\
&+\bigg\|\biggl( \frac{1}{N} \X'\X \biggr)\inv\bigg\| \bigg\|\frac{1}{N}\biggl(\Xii'\Mi\inv\Hij+\Xjj' \biggr)\bigg\| \label{eqn:betabij_b}\\
&\hspace{5mm} \bigg\|\bigl(\Mj-\Hij'\Mi\inv\Hij\bigr)\inv\bigg\| \label{eqn:betabij_c}\\
&\hspace{5mm} \bigg\|\Hij' \Mi\inv\resi+\resj\bigg\| \label{eqn:betabij_d}
\end{align}
\noindent where the left-hand-side is $o_p(1)$ following the reasoning used in~\eqref{eqn:bijb_a}-\eqref{eqn:bijb_f}. Thus, the difference $\bfe -\bij\overset{p}{\to} \zero$ as $N\to\infty$ and $T$ fixed.

\subsection{Conditional Influence}
\noindent Rearrange~\eqref{eqn:bijb}
\begin{equation}\label{eqn:bfe_beta_bji2}
\begin{split}
\small
\sqrt{N}&\bigl(\bij  - \bbeta \bigr)  - \sqrt{N}\bigl(\bfe  - \bbeta \bigr)  =  \\
& - \biggl( \frac{1}{N} \X'\X \biggr)\inv \biggl\{\frac{1}{\sqrt{N}}\, \Xii' \Mi\inv\resi 
 +\frac{1}{\sqrt{N}} \bigl(\Xii'\Mi\inv\Hij+\Xjj'\bigr) 
\bigl(\Mj-\Hij'\Mi\inv\Hij\bigr)\inv \bigl(\Hij' \Mi\inv\resi+\resj\bigr) \biggr\}
\end{split}
\end{equation}
\noindent and~\eqref{eqn:bib} replacing $i$ with $j$
\begin{equation}\label{eqn:bfe_beta_bi2}
	\sqrt{N}\bigl(\bj  - \bbeta \bigr) -\sqrt{N}\bigl(\bfe  - \bbeta \bigr) = - \biggl( \frac{1}{N} \X'\X \biggr)\inv \frac{1}{\sqrt{N}}\, \Xjj' \Mj\inv\resj.
\end{equation}
\noindent The difference of \eqref{eqn:bfe_beta_bji2} and \eqref{eqn:bfe_beta_bi2} is as follows
\begin{multline}
\sqrt{N}(\bij - \bbeta)  -\sqrt{N}(\bj- \bbeta) = \notag \\
   \biggl( \frac{1}{N} \X'\X \biggr)\inv  \biggl\{\frac{1}{\sqrt{N}}\, \Xjj' \Mj\inv\resj -\frac{1}{\sqrt{N}}\, \Xii' \Mi\inv\resi  \\
 -\frac{1}{\sqrt{N}} \bigl(\Xii'\Mi\inv\Hij+\Xjj'\bigr) 
\bigl(\Mj-\Hij'\Mi\inv\Hij\bigr)\inv \bigl(\Hij' \Mi\inv\resi+\resj\bigr) \biggr\}
\end{multline}
\noindent Then,  by the Reverse Triangle Inequality 
\begin{align}
\bb \sqrt{N}&(\bij   - \bbeta) -\sqrt{N}(\bj- \bbeta) \bb \notag \\
  & \le  \biggl\|\biggl( \frac{1}{N} \X'\X \biggr)\inv \biggl\|   \biggl\{ \biggl\|\frac{1}{\sqrt{N}}\, \Xjj' \Mj\inv\resj \biggl\| +  \biggl\|\frac{1}{\sqrt{N}}\, \Xii' \Mi\inv\resi  \biggl\| \\
 &+  \bigg\|\frac{1}{\sqrt{N}} \bigl(\Xii'\Mi\inv\Hij+\Xjj'\bigr) \bigg\| 
\bigg\|\bigl(\Mj-\Hij'\Mi\inv\Hij\bigr)\inv \bigg\| \bigg\| \bigl(\Hij' \Mi\inv\resi+\resj\bigr) \biggl\| \biggr\}
\end{align}
is  $o_p(1)$  by same arguments used for~\eqref{eqn:bbi} and~\eqref{eqn:bbij_a}-\eqref{eqn:bbij_d}. Therefore, 
\begin{equation}
 \sqrt{N}\big(\bij   - \bbeta\big) = \sqrt{N}\big(\bj- \bbeta\big) + o_p(1)
\end{equation}
\noindent which is asymptotically equivalent to the FE estimator $\bfe$ by result~\eqref{eqn:bi_convergence} relative to unit $j$.

\end{document}